\documentclass[aip,amsmath,amssymb,reprint]{revtex4-1}

\usepackage{dcolumn}
\usepackage{bm}
\usepackage{siunitx}
\usepackage{graphicx}
\usepackage{svg}
\usepackage{ulem}

\begin{document}



\title{Hollow-core fiber loading of nanoparticles into ultra-high vacuum} 

\author{Stefan Lindner}
\author{Paul Juschitz}
\author{Jakob Rieser}
\author{Yaakov Y. Fein}
\email[]{yaakov.fein@univie.ac.at}
\author{Mario A. Ciampini}
\affiliation{Faculty of Physics, Vienna Center for Quantum Science and Technology (VCQ), University of Vienna, Boltzmanngasse 5, A-1090 Vienna, Austria}
\author{Markus Aspelmeyer}
\affiliation{Faculty of Physics, Vienna Center for Quantum Science and Technology (VCQ), University of Vienna, Boltzmanngasse 5, A-1090 Vienna, Austria}
\affiliation{Institute for Quantum Optics and Quantum Information (IQOQI) Vienna, Austrian Academy of Sciences, A-1090 Vienna, Austria}
\author{Nikolai Kiesel}
\affiliation{Faculty of Physics, Vienna Center for Quantum Science and Technology (VCQ), University of Vienna, Boltzmanngasse 5, A-1090 Vienna, Austria}

\date{\today}

\begin{abstract}
Many experiments in the field of optical levitation with nanoparticles today are limited by the available technologies for particle loading. Here we introduce a new particle loading method that solves the main challenges, namely deterministic positioning of the particles and clean delivery at ultra-high vacuum levels as required for quantum experiments. We demonstrate the efficient loading, positioning, and repositioning of nanoparticles in the range of $100-755\,$\si{\nano\metre} diameter into different lattice sites of a standing wave optical trap, as well as direct loading of nanoparticles at an unprecedented pressure below $10^{-9}\,$\si{\milli\bar}. Our method relies on the transport of nanoparticles within a hollow-core photonic crystal fiber using an optical conveyor belt, which can be precisely positioned with respect to the target trap. Our work opens the path for increasing nanoparticle numbers in the study of multiparticle dynamics and high turn-around times for exploiting the quantum regime of levitated solids in ultra-high vacuum.


\end{abstract}


\maketitle

Optical levitation and motional control of dielectric objects in vacuum provide a unique platform for experiments in fundamental and applied research~\citep{levitoreview,Millen_2020}. 
Optically levitated nanoparticles have been used to study stochastic thermodynamics in underdamped environments~\citep{millen2018}, develop novel sensing schemes \citep{Ranjit2016,Hempston2017}, search for new physics \citep{Geraci2010,Moore20a,Aggarwal2022} and explore many-body phenomena ~\citep{Zemanek2008,Rieser2022,vijayan2023cavitymediated}. 
Owing to the flexible spatial and temporal control of the potential landscape and the excellent isolation from the environment~\citep{Bateman2014,Kaltenbaek2016,Neumeier2022}, levitated optomechanics is also a promising platform for probing macroscopic quantum mechanics or utilizing quantum enabled sensing applications.
Recent experiments have entered the quantum regime by preparing nanoparticles in 
their motional ground state~\citep{Delic2020,Magrini2021,tebbenjohanns2021a,Ranfagni2022,Kamba2022,Piotrowski2023}.
The next generation of quantum experiments will require extended coherence times and hence ultra-high vacuum (UHV) environments at or below $10^{-10}\,$\si{mbar} to avoid collisional decoherence~\citep{Bateman2014,Weiss2021,Neumeier2022}. 
Such vacuum levels per se are readily achieved in the laboratory. However, the available methods for nanoparticle loading to date make room-temperature optical levitation below $10^{-9}\,$\si{mbar} a challenging endeavor.

The most established method for loading nanoparticles into optical traps is based on ultrasonic nebulizers, which spray a dilute nanoparticle solution directly into the vacuum chamber at almost atmospheric pressure.
While reliable, disadvantages of this method are contamination of the vacuum system and the in-vacuum optics with solvent and nanoparticles. 
To reach UHV, nebulizer loading followed by a bakeout is in principle possible, but the long turnaround times together with potential misalignment of optics make this approach impractical in most cases. 
The method is also probabilistic, which is undesirable when working with multiple traps.

Various other approaches have been explored to address the challenge of loading particles into UHV environments. 
Loading from vibrating surfaces~\citep{Ashkin2003,Ayub2022,Weisman2022} has been demonstrated for particle sizes down to \SI{85}{\nano\metre} in diameter. 
This method eliminates solvent contamination, but it also ejects particles at high speeds and with a large angular spread, making efficient capture in UHV difficult. 
Another solvent-free loading technique is laser-induced acoustic desorption (LIAD) ~\citep{Asenbaum2013,Bykov2019,liadmillen}. 
As for piezo loading, however, the speed and emission angle of particles currently prepared with this method do not allow for direct deterministic loading into optical traps at UHV. 
Finally, particles have been transferred using load-lock schemes~\citep{Mestres2015,AndreaLoadLock} in a contamination-free and deterministic way. 
While this method has demonstrated transfers down to $6\times10^{-5}\,$\si{\milli\bar}, it still requires additional pump-down time to reach the desired UHV level.

\begin{figure*}[t]
    \centering
    $\vcenter{\hbox{\includegraphics[width=1.\textwidth]{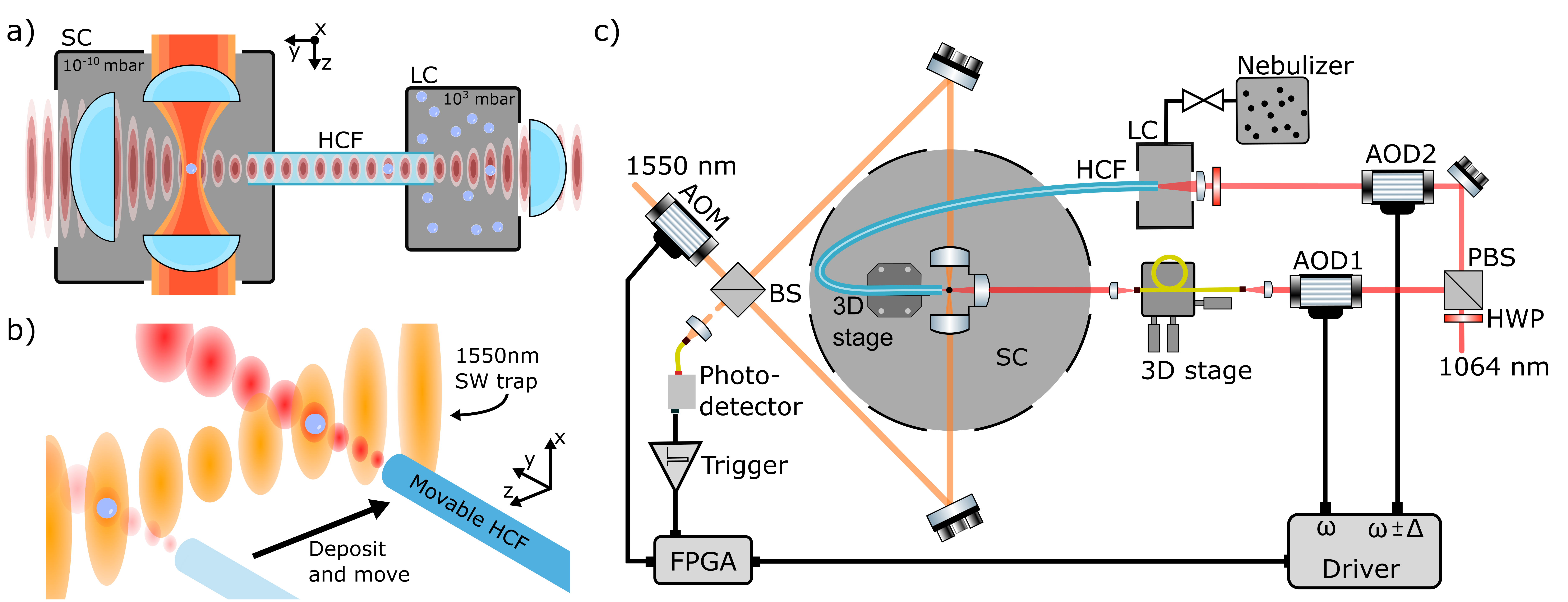}}}$
    \caption{a) An illustration of the HCF loading technique. A hollow-core photonic crystal fiber (HCF) connects the UHV ``science chamber" (SC) to a low vacuum ``loading chamber" (LC). An optical standing wave consisting of two counter-propagating lasers coupled into the HCF can, via detuning one arm, transport particles from the low vacuum chamber to the high vacuum region and deposit the particles directly into the target trap. b) The technique's capability to address multiple traps. In this case, different nodes of a standing wave are addressed by moving the HCF to the relevant positions. c) A schematic depiction of the \SI{1064}{nm} (red beam) optical conveyor belt, the science chamber, loading chamber and nebulizer, the \SI{1550}{nm} (orange beam) target trap in its Sagnac configuration and the particle detection in the dark port of the interferometer. The target trap intensity is switched via an acousto-optic modulator (AOM). The HCF standing wave can moved by applying a detuning to one of the acousto-optic deflectors (AOD). The 3D stage (Thorlabs NanoMax) is an auto-alignment system used to maintain coupling to the HCF as the fiber tip in the science chamber is translated.}
    \label{fig:schematic}
\end{figure*}

Here, we experimentally demonstrate loading of nanoparticles from atmospheric pressure directly into UHV levels below $10^{-9}\,\mathrm{mbar}$. Our method works on the timescale of minutes, without contamination of the UHV chamber, and is spatially precise, i.e. it delivers particles to specific locations. 
Our approach is based on an optical conveyor belt\citep{Meschede2001} inside a hollow-core photonic crystal fiber (HCF)~\citep{RussellHCF} for transporting sub-micron particles\citep{Grass2016}. 
Silica particles ranging from \SI{100}{\nano\metre} to \SI{755}{\nano\metre} in diameter are loaded into the optical conveyor belt via an ultrasonic nebulizer and are transported one by one from a low-vacuum loading chamber into a UHV science chamber (see Fig. \ref{fig:schematic}a). 
The fiber core diameter of \SI{9}{\micro \metre} and the fiber length of \SI{1.4}{\metre} enable pressure differentials between the chambers of up to 12 orders of magnitude~\citep{Ermolov2013,Abraham:98,supp}.
After transport, the nanoparticles are directly deposited into a standing-wave trap in the science chamber. 
The fiber tip can then be re-positioned and another particle loaded into a different trapping site if required (see Fig. \ref{fig:schematic}b). 
Loading directly into UHV is realized via an active trigger mechanism which enables capture of the particle in the target trap without the aid of gas damping.

Our experiment consists of an optical conveyor belt for particle transport from a high-pressure loading chamber to a target optical trap in a clean UHV chamber (Fig. \ref{fig:schematic}c). 
Various target trap configurations are compatible with our loading technique; the only requirement is the ability to trigger a rapid switching of the potential depth once the particle enters the trap.
In our case, we use a \SI{1550}{\nano\metre} standing wave trap in a Sagnac configuration with approximately $1\,$W input power. 
The power switching is provided via an AOM at the Sagnac input, while we perform motional readout of the particle via the dark port of the Sagnac interferometer.
We use a pair of 0.6$\,$-$\,$NA aspheric lenses near the center of the Sagnac to achieve both a tight trap focus and a large collection efficiency for readout.  

The conveyor belt used for transporting the particles inside the HCF is a moving optical standing wave formed by two \SI{1.5}{W} counter-propagating \SI{1064}{\nano\metre} beams (Azurlight ALS-IR-1064 10W), where one beam is coupled into the high-pressure side of the HCF and the other into the UHV side.
Each beam is frequency controlled via an AOD, which allows one to set a relative detuning $\Delta$ and thereby create a moving standing wave with velocity $v\,=\,\lambda\Delta/2$. 
The target-trap-facing end of the HCF is mounted on a UHV-compatible 3D-translation stage (Attocube ANPx101/UHV) to enable precise positioning of the particles. 
The ability to translate the HCF inside the science chamber is useful for trap alignment as well as for loading multiple particles.
To maintain the standing wave quality during translation of the HCF, we actively stabilize the coupling of the counter-propagating beam into the HCF using a 3D-piezo positioner (Fig. \ref{fig:schematic}c).

\begin{figure}[h]
    \centering
    \includegraphics[width = 1\columnwidth]{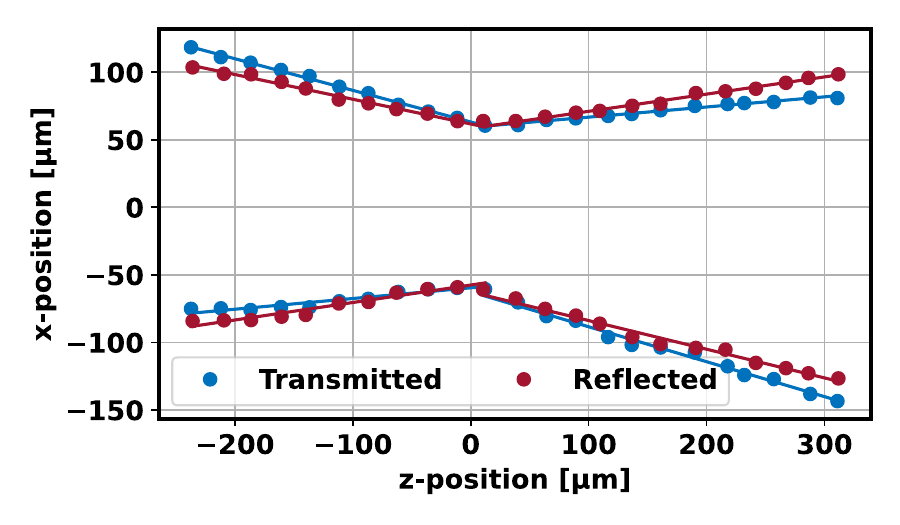}
    \caption{Determination of the target trap geometry via a knife-edge style measurement. We block part of the target trap beam with the HCF tip and measure the $x$ positions at which the transmitted power is reduced to $(1-1/e^2)$ of its original value. By performing this measurement at various axial ($z$) positions we can reconstruct the profiles of both beams comprising the standing wave target trap (blue and red dots) and determine the focal positions with respect to the HCF tip. This allows us to align the HCF to the deepest part of the target trap potential.}
    \label{fig:grazing}
\end{figure}

To align the HCF optical conveyor belt to the target trap, the HCF tip is used to map the target trap geometry~\citep{supp}. 
This allows us to determine the positions of the beam foci of the standing wave target trap with respect to the HCF as well as their respective waists (see Fig. \ref{fig:grazing}). 
We then position the HCF such that particles are deposited into the center of the target trap, where the trap is deepest.

\begin{figure}[h!]
    \centering
    \includegraphics[width=0.9\columnwidth]{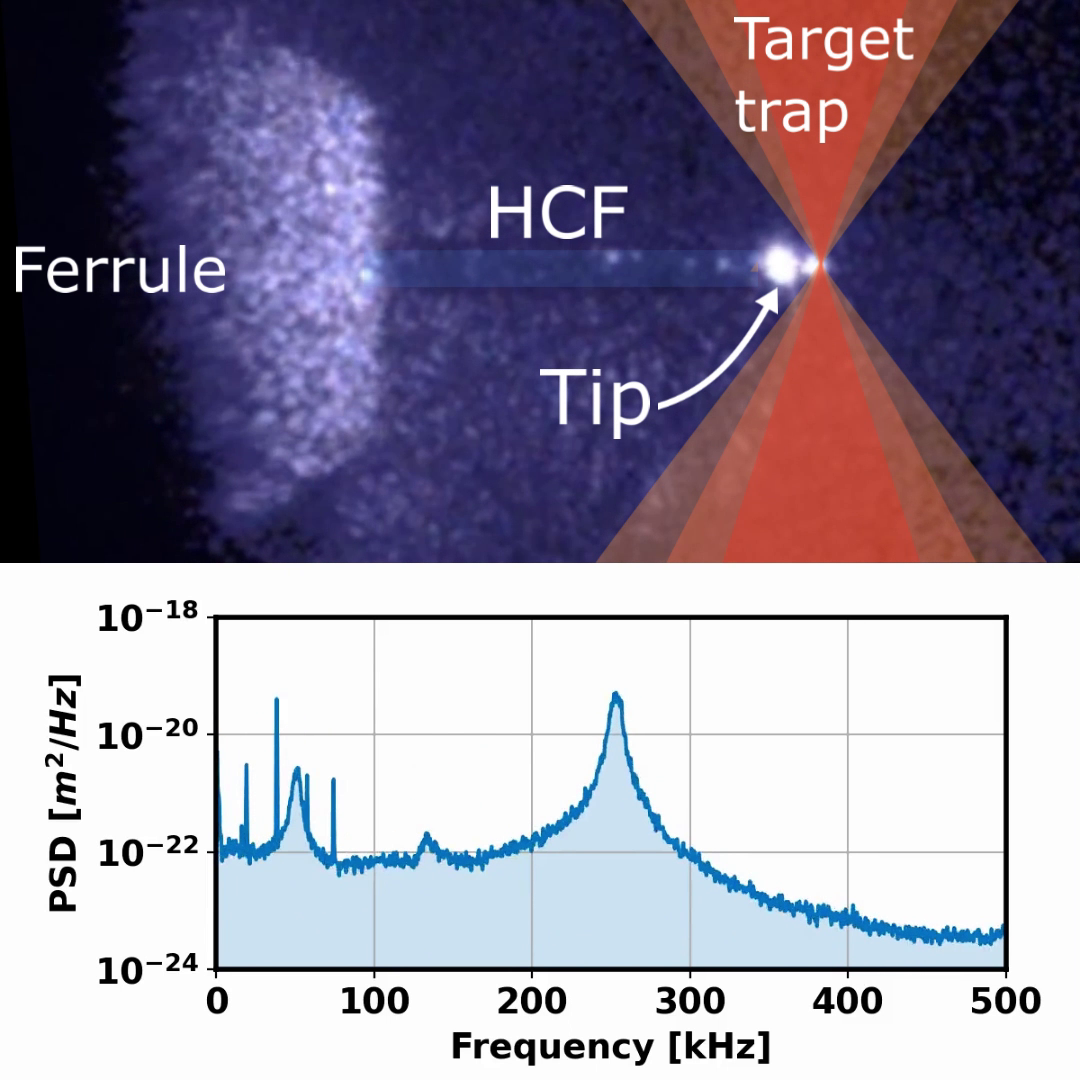}
    \caption{Top: Video sequence of a complete transfer of a \SI{100}{\nano\metre} diameter particle from HCF conveyor belt to our target trap (at total power of $\sim \,$ \SI{1}{W}) at a pressure of \SI{8}{\milli\bar}. The overlays in the upper part show the ceramic ferrule in which the HCF is mounted, the HCF, the fiber tip (bright due to scattering), and the target trap. The video goes dark at the end as the conveyor belt is turned down since the recording camera is not sensitive to the \SI{1550}{\nano\metre} of the target trap. Bottom: The corresponding power spectral density of the signal detected in the dark port of the Sagnac target trap (multimedia available online\citep{vid}).}
    \label{fig:manual_handover}
\end{figure}

The procedure for loading nanoparticles via the HCF into the target trap is as follows:
Nanoparticles are diluted in isopropanol and loaded via a nebulizer into the loading chamber at $\sim$\SI{1}{\bar}.   
A particle falling into one of these trap sites (as confirmed by an infrared camera) is transported along the moving antinodes from the loading chamber to the science chamber.
We first describe the high-pressure case ($>\,1\,\mathrm{mbar}$ in the science chamber), where the particle can be stably levitated in front of the HCF.
When the particle arrives near the exit of the HCF, the detuning is reduced to decrease the particle velocity for its transfer to the target trap (see Fig. \ref{fig:manual_handover} (multimedia available online)). 
As the particle enters the target trapping region its mechanical motion becomes visible in the dark-port detection signal.
At the trap center (where the trapping frequencies are maximized), the conveyor belt is turned off, thereby completing the transfer.
Fig. \ref{fig:manual_handover} (multimedia available online~\citep{vid}) shows a video sequence of a complete transfer of a \SI{100}{\nano\metre} diameter particle at a pressure of \SI{8}{\milli\bar} together with the power spectral density of the dark-port signal recorded during this procedure.
Particle transfers into a UHV environment require an additional trigger mechanism, as described below.

\begin{figure}[h]
    \centering
    \includegraphics[width = 1.\columnwidth]{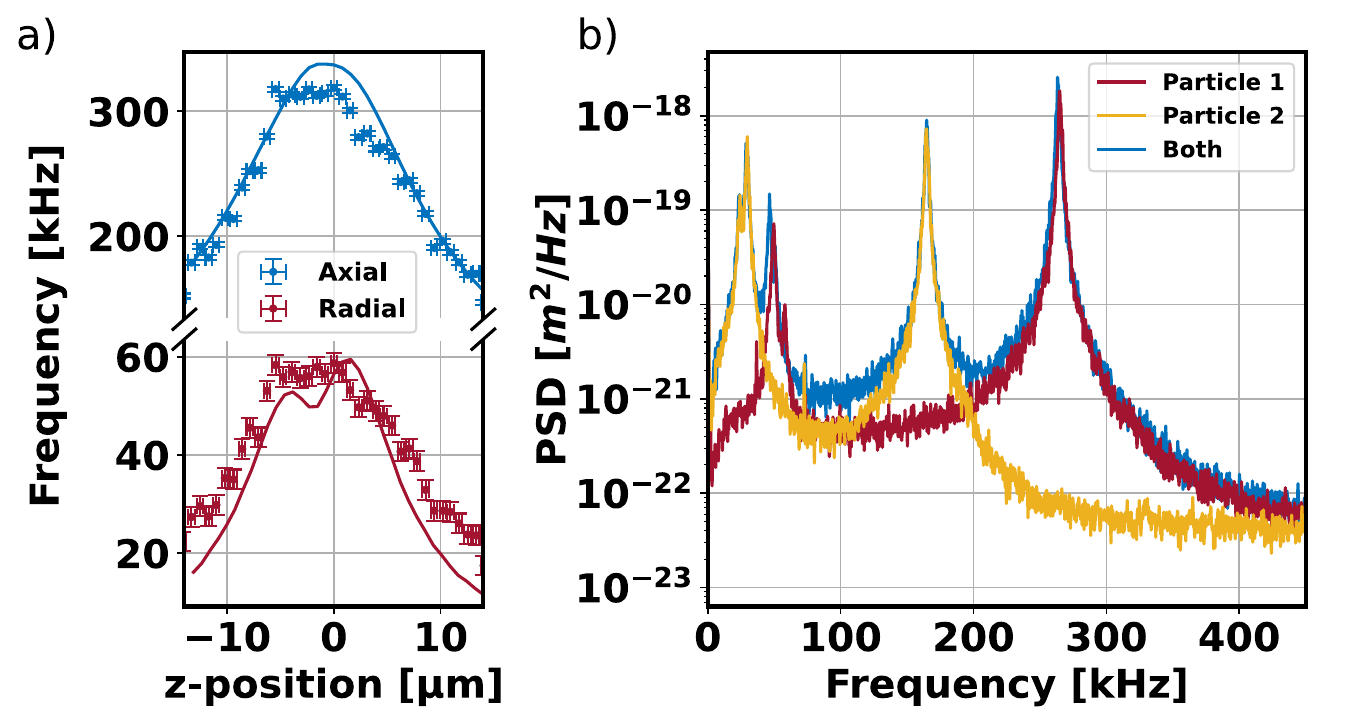}
    \caption{a) Demonstration of the spatial resolution of the HCF loading technique. We deposit a particle from the HCF conveyor belt to our target trap at a pressure of \SI{5}{\milli\bar}, turn off the conveyor belt and measure the frequency in the Sagnac trap. Then the conveyor belt is reactivated, the particle pulled back into the fiber, and then placed into the next antinode of the target trap. This procedure is repeated over a range of $\sim\,$\SI{30}{\micro \metre}. Shown are the measured axial (radial) frequencies in blue (red) at each position and the theoretically expected frequencies\citep{supp}. b) Power spectral densities demonstrating controlled loading and removal into/from various trap positions with two particles. Particle 1 (red spectrum) is loaded via the HCF into an antinode close to the trap focus, then the HCF is moved to a position seven antinodes away, and Particle 2 is loaded simultaneously (blue spectrum). Finally, the HCF is moved back to its original position, and Particle 1 is removed, leaving only Particle 2 (yellow spectrum).}
    \label{fig:antinodes}
\end{figure}

A significant benefit of the HCF-loading technique is the capability to deterministically deliver and retrieve particles with micron-level spatial resolution. 
We demonstrate this by loading and manipulating several particles controllably into different standing wave trap sites at \SI{5}{\milli\bar} as shown schematically in Fig. \ref{fig:schematic}b.
Specifically, we show how a single particle can be deposited, retrieved and redeposited into successive antinodes of our target trap. The measured frequencies are in agreement with a theoretical model determined by the measured trap geometry (Fig. \ref{fig:antinodes}a).
Discrepancies with the model predictions are attributed to an observed slipping behavior of the particle into a neighboring antinode, likely due to the tilt of the HCF with respect to the target trap axis.
Finally, we demonstrate how multiple particles can be loaded into desired trap sites simultaneously and also retrieved back into the HCF (see Fig. \ref{fig:antinodes}b). 

\begin{figure}[h]
    \centering
    \includegraphics[width = 0.94\columnwidth]{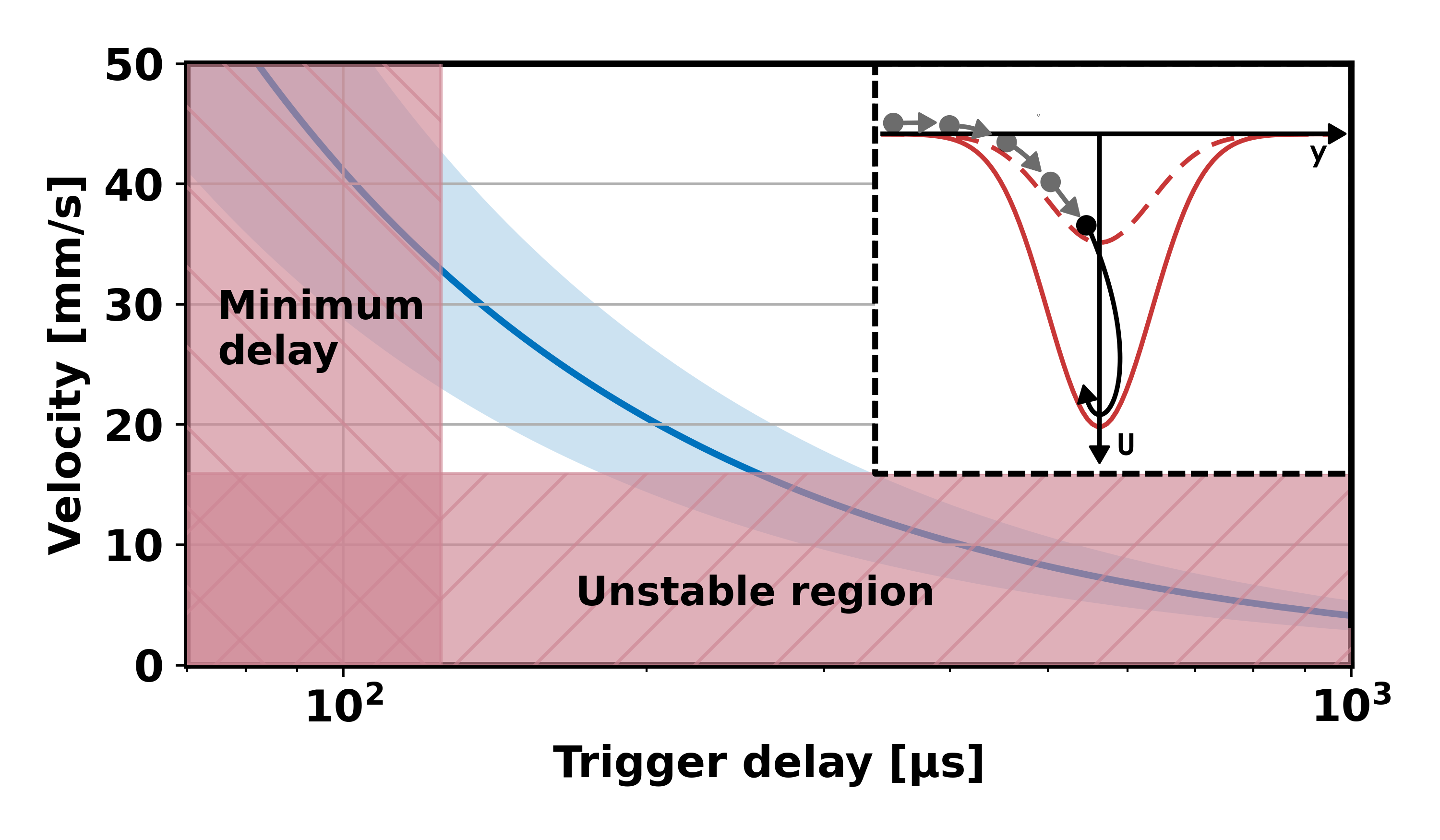}
    \caption{
    Optimal velocity of the particle in the optical conveyor belt for trap transfers. The blue line gives the particle velocity upon exiting the HCF which corresponds to the particle traveling to the center of the target trap over the time it takes for the trigger to activate. Velocities close to this blue curve are desirable since then the particle will be near the trap center when the trap depth is increased, making a successful transfer more likely. The shaded region displays a 25\% error on the optimal velocity which we found allowed for reliable trap transfers. The red regions correspond to empirical bounds on the velocity and trigger delay respectively as discussed in the main text. The inset sketches a particle trajectory as the trigger activates and deepens the trapping potential (dashed to solid red).}
    \label{fig:trigger}
\end{figure}

The particle transfer procedure described above entails depositing particles with slow detuning into the target trap. 
This requires stable levitation in front of the hollow-core fiber, which we have observed to work reliably only down to approximately \SI{1}{\milli\bar} (without feedback).
To load particles into UHV, we instead ballistically eject the particle from the HCF at a constant detuning toward the target trap.
Without gas damping, the conservative nature of the trapping potential means that the particle cannot be trapped without additional deceleration.
Therefore, to transfer particles directly into a UHV environment we employ a trigger that deepens the target trap as the nanoparticle passes its center~\citep{Bykov2019} (see inset Fig. \ref{fig:trigger}). 

We monitor the particle signal in the dark port of the Sagnac interferometer to identify the moment the particle enters the target trap.
The resulting detector signal is sent to a three-stage trigger circuit consisting of an amplifier, rectifier and comparator. 
If the signal exceeds the comparator threshold, the trigger sends a TTL pulse to an FPGA~\citep{supp} which, after a certain delay, simultaneously increases the target trap depth and turns off one conveyor belt beam. 
By leaving the counter-propagating HCF beam on, the nanoparticle is additionally decelerated by radiation pressure which assists the loading process.


For the trigger to work effectively, it needs to increase the target trap power when the particle is near the trap center. 
There is an experimental delay between the detection of the particle in the target trap and the moment when the trap power is increased.
This delay is at least \SI{125}{\micro \second}, but in practice will be longer depending on the specific settings of the trigger circuit~\citep{supp}.
To make the time the particle crosses the trap center coincide with the trap activation, we choose the particle velocity accordingly (via the conveyor belt velocity).
The minimum delay sets an upper limit on the particle velocity  (see Fig. \ref{fig:trigger}):
above \SI{33}{\milli\metre/\second}, the particle will have passed the trap center before the potential deepens.
A lower limit is given by the stability of the conveyor belt trap:
we observed that at science chamber pressures below $\sim 10^{-2}$ \si{\milli\bar}, particles moving slower than \SI{16}{\milli\metre/\second} escape the conveyor belt before reaching the target trap.
So, quickly traversing the low-pressure region avoids particle loss. Alternatively, the conveyor belt trap stability could be improved by active feedback cooling.

\begin{figure}[h]
    \centering
    \includegraphics[width = 1.\columnwidth]{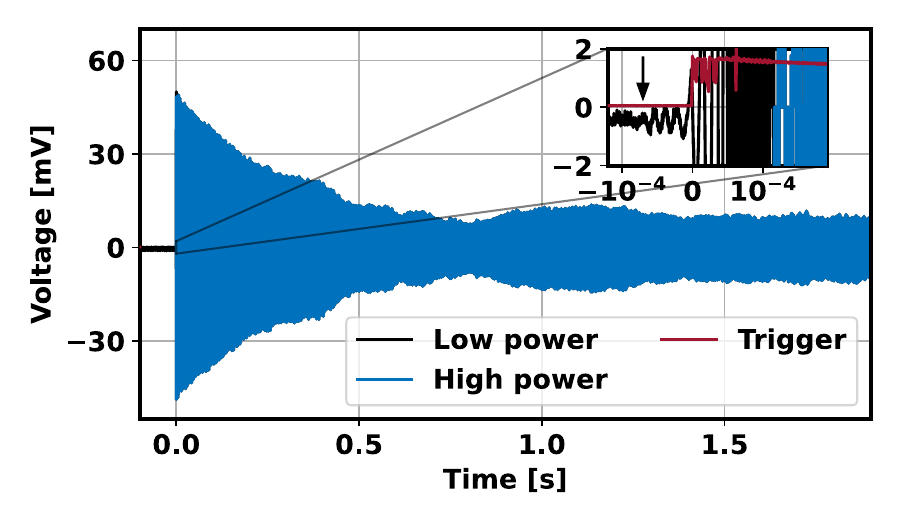}
    \caption{Timetrace of a \SI{365}{nm} particle being transferred from the HCF conveyor belt into the target trap at $2\times 10^{-9} \si{mbar}$ as measured in the dark port of the Sagnac target trap. The inset shows the particle entering the trap (black arrow), the trigger output once the increase in variance is detected (red) and the resulting power switching (black to blue). The total trigger delay (the time from initial particle detection in the target trap to the trap depth increase) in this case is about \SI{185}{\micro \second}.}
    \label{fig:timetrace}
\end{figure}

The procedure for UHV transfers is then similar to that of the high-pressure case described above, except for a higher particle velocity when leaving the HCF and the addition of the triggering mechanism. 
Once the trigger increases the target trap power and turns off the conveyor belt beam, the particle is fully trapped and the remaining conveyor belt beam can be shut off.
As an example, Fig. \ref{fig:timetrace} shows successful loading of a \SI{365}{\nano\metre} silica nanoparticle at a science-chamber pressure of $2\times10^{-9}\,\mathrm{mbar}$.
We used this method to load particles in a diameter range between \SI{143}{\nano\metre} and \SI{365}{\nano\metre} into a standing wave trap at UHV. 
We have successfully demonstrated transfers down to a pressure of $8\times10^{-10}\, \si{mbar}$, limited only by the loading chamber pressure during these experimental runs\citep{supp}.
A complete handover from start (particle trapped in conveyor belt) to end (particle in the target trap at UHV) takes less than three minutes, even though the trigger setup is controlled manually. We expect that automation could reduce the loading time to about ten seconds.

Our hollow-core fiber loading technique enables fast, clean and deterministic delivery of nanoparticles to optical traps in ultra-high vacuum. 
It can also be applied to different trapping schemes and particle sizes, and is particularly suited for compact on-chip levitation platforms. 
If desired, one could also use other schemes such as piezo loading or LIAD to load the HCF in lieu of a nebulizer.

Our technique opens the door to room-temperature levitated optomechanics experiments at unprecedented vacuum levels, a prerequisite for a number of proposed experiments aiming to utilize the quantum regime of levitated solids for both fundamental science and sensing applications. 
As this method enables high-resolution positioning of nanoparticles, it is of interest to the growing community exploring architectures with multiple traps.

\begin{acknowledgments}
The authors would like to thank Oliver Gabriel for the design of the electronic trigger circuit and Gregor Meier for help with its implementation. This project was supported by the European Research Council under the European Union’s Horizon 2020 research and innovation program (ERC Synergy QXtreme, Grant No. 951234) and the Austrian Science Fund (FWF, Grant No. Y 952-N36 START).

\end{acknowledgments}

\bibliography{hcf_ref}

\end{document}


\title{Supplemental Material: Hollow-core fiber loading of nanoparticles into ultra-high vacuum} 

\author{Stefan Lindner}
\author{Paul Juschitz}
\author{Jakob Rieser}
\author{Yaakov Y. Fein}
\email[]{yaakov.fein@univie.ac.at}
\author{Mario A. Ciampini}
\affiliation{Faculty of Physics, Vienna Center for Quantum Science and Technology (VCQ), University of Vienna, Boltzmanngasse 5, A-1090 Vienna, Austria}
\author{Markus Aspelmeyer}
\affiliation{Faculty of Physics, Vienna Center for Quantum Science and Technology (VCQ), University of Vienna, Boltzmanngasse 5, A-1090 Vienna, Austria}
\affiliation{Institute for Quantum Optics and Quantum Information, Austrian Academy of Sciences, A-1090 Vienna, Austria}
\author{Nikolai Kiesel}
\affiliation{Faculty of Physics, Vienna Center for Quantum Science and Technology (VCQ), University of Vienna, Boltzmanngasse 5, A-1090 Vienna, Austria}

\date{\today}

\maketitle

\section*{Particle identification}
\begin{figure}[h!]
    \centering
    \includegraphics[width = 1\columnwidth]{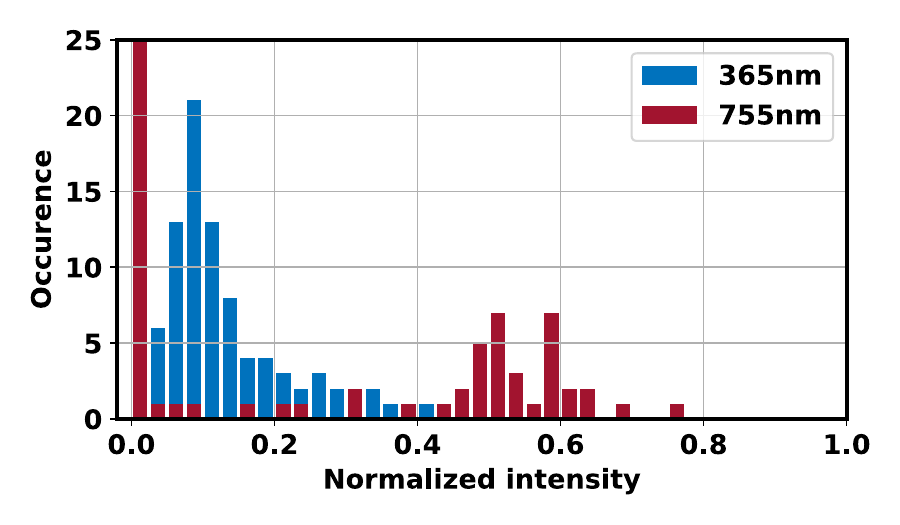}
    \caption{A brightness histogram that provides a simple method to differentiate silica particles from the smaller and continuously distributed isopropanol droplets as well as larger particle clusters. Camera settings, position of the particle in the fiber and polarization of the conveyor belt beam were kept constant throughout the measurement. }
    \label{fig:histogram}
\end{figure}

Since isopropanol droplets and clusters of particles rather than single particles are also regularly loaded into the HCF, it is useful to have a method to quickly differentiate these different species in situ.
We do this by analyzing scattering patterns in the HCF immediately after the loading chamber. 
In Fig. \ref{fig:histogram}, we show a brightness histogram (obtained by summing pixel values above the dark threshold) of a series of trapped objects in the HCF when loading with two solutions of mono-disperse particles. 
The discernible peaks in brightness allow us to distinguish individual particles from isopropanol droplets or particle clusters based on brightness characterization alone. 

\section*{Hollow-core fiber conductance}
\begin{figure}[h!]
    \centering
    \includegraphics[width = 1\columnwidth]{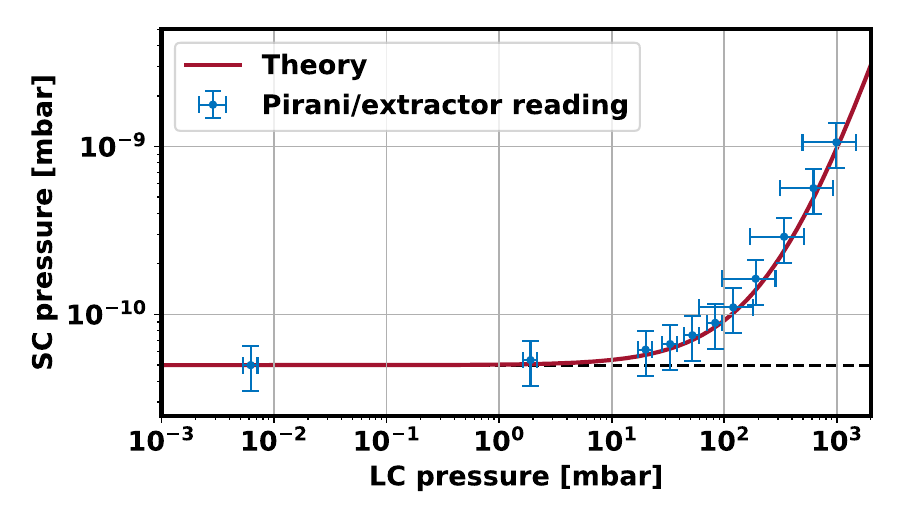}
    \caption{Vacuum conductance of the HCF. The science chamber (SC) pressure is shown as a function of the pressure in the loading chamber (LC). The lower asymptote defines the science chamber base pressure of $5\times10^{-11}$\si{\milli\bar}. The pressure rises to $1\times10^{-9}$ \si{\milli\bar} when the loading chamber is held at atmospheric pressure.}
    \label{fig:conductance}
\end{figure}

During loading, the pressure in the science chamber rises due to the small leak through the HCF from the loading chamber which is typically held at atmospheric pressure. 
The vacuum conductance through a \SI{1.4}{\metre} HCF was measured by varying the loading chamber pressure with nitrogen and measuring the corresponding pressure change in the science chamber, waiting at each point for the science chamber pressure to equilibrate.
The results are shown in Fig. \ref{fig:conductance}. 
The conductance depends on the HCF geometry (both the core and the smaller gaps of the photonic crystal), which we characterize with an image of a cleaved fiber. 
Significantly, it takes only $\sim\,$30 minutes to go from the $1 \times10^{-9}\,\mathrm{mbar}$ with the loading chamber at atmospheric pressure to $2\times 10^{-10}\,\mathrm{mbar}$. 
Therefore, applications requiring pressures below $10^{-10}\,$\si{mbar} would still benefit from this loading method, despite the small leak introduced by the HCF. 
Notably, this leak can be further decreased through a longer fiber or a lower loading pressure in the loading chamber.

\section*{Detailed alignment procedure}
We describe the geometric alignment procedure for overlapping the conveyor belt trap and the target trap in more detail. 
With one arm of the Sagnac trap blocked, we monitor the power $P_0$ transmitted through the science chamber as the tip of the HCF is used to partially obscure the open beam in a knife-edge style measurement. 
We determine the vertical beam extent at a given axial position by inserting the fiber into the beam from above (and below) and reading out the $x$-coordinates with the Attocube stage encoder at which the transmitted power decreases to $P_0\cdot\left(1-1/e^2\right)$. 
This procedure is repeated over a $z$-range of $\sim\,$\SI{0.5}{\milli\metre} in \SI{25}{\micro\metre} steps. 
A typical measurement is shown in the main text Fig. 2, where we have repeated the measurement for both beams of the Saganc. 
The longitudinal position of the beam focus can be found as the intersection along $z$ of two lines fitted to this measurement. 
The beam center along $x$ at the focal position can be found by taking the average $x$-position at this intersection point. 

By repeating this measurement for the previously blocked beam, one can deduce the relative axial and radial offsets of the foci of the two counter-propagating \SI{1550}{\nano\metre} beams. 
For $z \ll z_R$, the slope of the grazing positions $d\omega_{g}/dz$ determines the trap waist $\omega_0 \approx C\lambda/\pi(d\omega_{g}/dz)$ where $C$ is a factor determined by the chosen grazing position (for a $1/e^2$ power reduction, $C\approx 0.55$).
These waists and their respective offsets are used to obtain the theory curves for the expected trap frequencies of our standing wave target trap in the main text Fig 4a.

\section*{Refined alignment procedure for small trapping volumes}
The above method for overlapping the optical conveyor belt with the target trap works well in our case of a standing-wave target trap. 
This severely relaxes our axial alignment criterion as there is a wide range of stable trapping centers. 
When loading into a trap with a significantly smaller trapping volume, e.g., a high-NA single-sided tweezer, a more refined method may be beneficial. 
While loading from an HCF into a single-sided tweezer has already been demonstrated~\citep{Jakobthes}, the alignment method used there is incompatible with UHV systems. 

Here we outline a UHV-compatible technique that we tested in our setup.
For this method, one additionally requires motional detection of the nanoparticle inside the HCF~\citep{Grass2016}. 
The alignment procedure from the main text is used as an initial alignment step to narrow down the relevant search area. 
The target trap is then modulated at a frequency $\Omega_\mathrm{mod}$ via an AOM or EOM as the particle is slowly sent out of the HCF tip into the trapping area. 
The trap center is then determined as the position at which the harmonically-modulated gradient force, measured in the HCF detection, is maximized.
Note that additional stabilization may be required at UHV pressures to keep the particle in the conveyor belt trap during this procedure.

\section*{Detailed description of the trigger mechanism}
As described in the main text, we require a triggered trap activation for UHV trap transfers.
We use a three-stage analog circuit (amplifier, rectifier and comparator) to determine the moment the particle enters the edge of the target trap.
As the input to this circuit we use the signal from a photodetector in the dark port of our target trap Sagnac configuration (see main text Fig. 1c).
If there is no particle in the target trap, the variance of the input signal is small and stays below the threshold voltage of the comparator stage of the circuit. 
When a particle enters the target trap it scatters light into the dark port and the variance of the detected signal will exceed the comparator threshold, resulting in the output of a high TTL signal.

\begin{figure}[b!]
    \centering
    \includegraphics[width=\columnwidth]{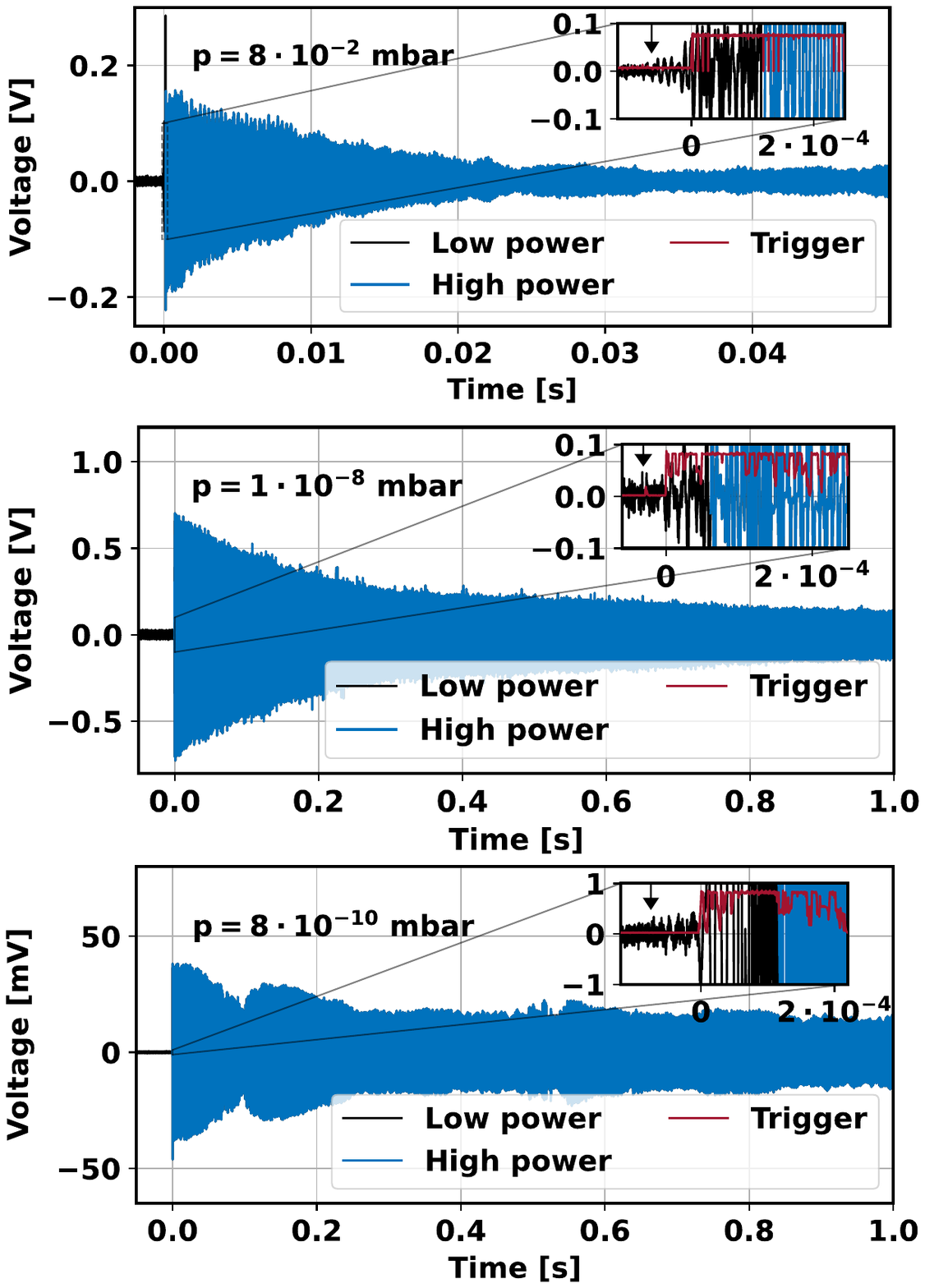}
    \caption{Three additional transfer traces at different pressures. As in the main text, the inset shows the moment of trap entry (marked through a black arrow).}
    \label{fig:si_handover}
\end{figure}

This TTL pulse is sent to our FPGA-based experimental control unit (Sinara) which triggers two simultaneous events. 
The Sinara RF driver ('Urukul"-module) increases the RF power sent to the AOM in the target trap arm and thus the target trap depth.
In parallel, the Sinara DAC (''Fastino"-module) sends a DC signal to a modulation input of our RF driver controlling the two AODs for the optical conveyor belt, turning off one of the arms.

\section*{Trigger delays}
The signal delay from the trigger circuit alone (between particle detection in the target trap and TTL output) depends on the comparator threshold setting but can be as low as \SI{10}{\micro\second}.
The main source of the ''minimum delay" region in main text Fig. 5 instead comes from the programming of the Sinara.
Using the ARTIQ environment~\citep{artiq}, we add a \SI{110}{\micro\second} delay between the receipt of the TTL pulse and the Sinara outputs to avoid underflow errors in the real-time I/O design; we typically observe \SI{115}{\micro\second} from the trigger TTL to the power switching of the target trap.
Taken together, the minimum total delay from the moment the particle is detected in the target trap until the moment the trigger activates the power switching is at least \SI{125}{\micro\second}.
The actual delay is typically larger and depends on the chosen comparator threshold, which is a trade-off for the sensitivity of the trigger circuit and the number of false trigger events. 

\section*{Transfers at different pressures}
Fig. \ref{fig:si_handover} shows additional trap transfers at a range of different pressures, including the to-date lowest pressure ($8\times10^{-10}\, \si{mbar}$) achieved for a direct transfer.

\bibliography{hcf_ref}